\documentclass[12pt]{article}
\usepackage{latexsym}
\usepackage{amssymb}
\usepackage{amsmath}
\usepackage{graphicx}

\newcommand{\V}{V(\phi)}
\newcommand{\vprime}{V^{\prime}(\phi)}
\newtheorem{theorem}{Theorem}[section]
\newtheorem{lemma}[theorem]{Lemma}

\newcommand{\tach}{\Phi}
\newcommand{\squart}{\sqrt{1+\tach^{;\kappa}\tach_{,\kappa}}}
\newcommand{\overbar}{\overline}

\begin{document}
\title{The gravitating perfect fluid-scalar field equations: quintessence and tachyonic}
\author {{\small A. DeBenedictis \footnote{e-mail: adebened@sfu.ca}} \\
\it{\small Department of Physics} \\ \it{\small Simon Fraser
University, Burnaby, British Columbia, Canada V5A 1S6}
 \and 
{\small A. Das \footnote{e-mail: das@sfu.ca}} \\
\it{\small Department of Mathematics} \\ \it{\small Simon Fraser
University, Burnaby, British Columbia, Canada V5A 1S6} \\
 \and
{\small S. Kloster \footnote{e-mail: stevek@sfu.ca}} \\
\it{\small Centre for Experimental and Constructive Mathematics}\\ \it{\small Simon Fraser
University, Burnaby, British Columbia, Canada V5A 1S6}}

\date{{\small February 5, 2004}}
\maketitle

\begin{abstract}
\noindent The system consisting of a self gravitating perfect fluid and scalar field is considered in detail. The scalar fields considered are the quintessence and ``tachyonic'' forms which have important application in cosmology. Mathematical properties of the general system of equations are studied including the algebraic and differential identities as well as the eigenvalue structure. The Cauchy problem for both quintessence and the tachyon is presented. We discuss the initial constraint equations which must be satisfied by the initial data. A Cauchy evolution scheme is presented in the form of a Taylor series about the Cauchy surface. Finally, a simple numerical example is provided to illustrate this scheme. 
\end{abstract}

\vspace{3mm}

\section{Introduction}
In the arena of cosmology and stellar structure, the importance of the perfect fluid source cannot be understated. The assumption of large-scale isotropy will demand, via the field equations, that the stress-energy tensor supporting the universe possess the algebraic 
structure of a perfect fluid. As well, cosmologies containing scalar fields have long been considered. The motivation being that the scalar field would represent some exotic component of matter, which could explain various puzzling phenomena presented by observational cosmology. More recently, viable alternatives to the standard big bang theory have been put forward, the most promising of which involve an inflationary era at some early time. Most of these theories invoke a scalar field to play the role of the inflaton in the early universe. At late times, the scalar field may again play an important role as a ``dark energy'' field. It has been observed that the universe has recently entered an acceleration phase and some exotic dark energy must presently dominate \cite{ref:accel1}, \cite{ref:accel2}. It is well known that a scalar field provides a simple model to explain this acceleration.

\qquad The scalar field may take many forms. Aside from the traditional quintessence field, which arises from a Lagrangian motivated by relativistic continuum mechanics, there is also the recently popular ``tachyonic'' scalar field from string theory, which possesses a Born-Infeld type action. Both of these fields have been used extensively in cosmology as the inflaton, dark matter and dark-energy (see \cite{ref:quint1}, \cite{ref:quint2}, \cite{ref:tach0}, \cite{ref:tach1}, \cite{ref:tach2}, \cite{ref:tach3}, \cite{ref:tach3point5}, \cite{ref:tach3point51}, \cite{ref:tach4}, \cite{ref:tach5}, \cite{ref:tach5a}, \cite{ref:tach5b}, \cite{ref:tach6}, \cite{ref:tach7}, \cite{ref:tach8} \cite{ref:tach8point5} and references therein). Usual cosmological models are spherically symmetric. However, more general cosmological models, admitting a three-parameter group of motions (Bianchi type I-IX) have also been studied.

\qquad For the above reasons, we believe that it is important to study the combined system of Einstein-scalar field-perfect fluid equations in detail. Specifically, in section 2 we present the system in its full generality for both the quintessence field and tachyon field. We discuss the system of equations and the number of functions which can be apriori prescribed. This is dictated by the number of functions present as well as the number of equations and identities. The eigenvector and eigenvalue structure is briefly discussed as well as its relation to physical quantities such as energy density and pressure.

\qquad The Cauchy problem has long been studied in general relativity and is of great use due to the difficulty involved in obtaining exact solutions to physical problems within the theory (see \cite{ref:lichbook}, \cite{ref:syngebook}, as well as \cite{ref:dascauch} where the gravitating complex scalar-Maxwell field system has been studied). In solving Einstein's equations, one often considers physically acceptable initial data (for example: data from plasma physics for stellar collapse, a specific form for the scalar potential inspired by string theory or particle physics in cosmology, etc.) The future evolution of this prescribed data is to be determined, ideally by analytically solving the Einstein equations and matter field equations of motion. However, this is often not possible and either some simplifying assumptions must be made or else approximation methods must be utilized. It is useful therefore to numerically evolve the initial data of the full, unsimplified model to determine the behaviour of the system at later times and, in some cases, to aid in determining what form the exact solution should take. A thorough treatment of the system consisting of a massless self-gravitating scalar field in the context of gravitational collapse (in the retarded time gauge) may be found in \cite{ref:christocauch}.

\qquad The Cauchy problem in general relativity is non-trivial. Initial data is not freely prescribable due to the existence of four constraining equations for the initial data. (These arise due to the differential identities among the system of field equations.) The constraint equations for both the quintessence plus perfect fluid and tachyon plus perfect fluid systems are derived in this paper along with the functions which one, in principle, may prescribe. In section 3 the Cauchy problem for the Einstein-perfect fluid-scalar field is properly posed using geodesic normal coordinates (which are particularly well suited to cosmological studies).
A scheme is presented which will evolve the this data from the initial $t=t_{0}=\mbox{constant}$ hyper-surface analytically into the bulk ($t > t_{0}$). Finally, we end with a simple example to illustrate the use of the scheme and its convergence properties. For an excellent review of the Cauchy problem in general relativity, the reader is referred to \cite{ref:friedrich}.

\section{General theory}
\subsection{Quintessence field}
The system considered consists of gravity coupled to a perfect fluid and quintessence scalar field, $\phi$. The fluid energy density, pressure and scalar potential are denoted by $\rho$, $p$ and $\V$ respectively. Derivatives with respect to the scalar field are denoted by a prime.

\qquad The quintessence Lagrangian density has the form \footnote{Conventions here follow that of \cite{ref:MTW}}:
\begin{equation}
\mathcal{L}_{q}=-\sqrt{g} \left[\frac{1}{2} \phi_{,\mu} \phi^{;\mu} +V(\phi) \right].
\end{equation} 

\qquad The Einstein and supplemental equations for this system are given by:
\begin{subequations}
\begin{align}
&\mathcal{E}^{\mu}_{\;\nu}:=G^{\mu}_{\;\nu}-8\pi T^{\mu}_{\;\nu}=0, \label{eq:einstQ} \\
&T^{\mu}_{\;\nu}:=\phi^{;\mu}\phi_{,\nu}-\left[\frac{1}{2} \phi_{,\alpha} \phi^{;\alpha}
+
V(\phi) \right] \delta^{\mu}_{\;\nu} + \left(\rho + p \right) u^{\mu}u_{\nu} +
p\delta^{\mu}_{\;\nu}, \label{eq:TQ} \\
&\mathcal{T}_{\nu}:=T^{\mu}_{\;\nu ;\mu}=0, \label{eq:consQ} \\
&\sigma:=\phi^{;\mu}_{\; ;\mu}-\vprime=0, \label{eq:SiQ} \\
&\mathcal{U}:=u^{\alpha}u_{\alpha}+1=0, \label{eq:UQ} \\
&\mathcal{K}:=\left[(\rho + p)u^{\alpha}\right]_{;\alpha} -u^{\alpha}p_{,\alpha}=0,
\label{eq:KQ} \\
&\mathcal{F}_{\nu}:=(\rho +p)u^{\alpha}u_{\nu;\alpha} +\left(\delta^{\alpha}_{\;\nu}
+u^{\alpha}u_{\nu}\right)p_{,\alpha}=0, \label{eq:FQ} \\
&\mathcal{C}^{\nu}(g_{\alpha\beta}, g_{\alpha\beta,\gamma})=0. \label{eq:CQ}
\end{align}
\end{subequations}
(The $\mathcal{C}^{\nu}$ denote four possible coordinate conditions.)
The algebraic and differential identities are:
\begin{subequations}
\begin{align}
&\mathcal{E}^{\alpha}_{\;\beta ;\alpha} +8\pi\mathcal{T}_{\beta} \equiv 0, \label{eq:id1Q}
\\
&\mathcal{T}_{\nu}- \left(\sigma \phi_{,\nu}+u_{\nu}\mathcal{K} +\mathcal{F}_{\nu}
\right)\equiv 0, \label{eq:id2Q} \\
&\frac{1}{2}\left(\rho+ p\right)u^{\nu}\left( \mathcal{U}-1\right)_{,\nu}
+\mathcal{U}u^{\nu}p_{,\nu} -u^{\nu}\mathcal{F}_{\nu} \equiv 0. \label{eq:id3Q}
\end{align}
\end{subequations}
In this system of equations the number of unknown functions is
\begin{equation}
10\left(g_{\alpha\beta}\right)+1\left(\phi\right)+1\left(V(\phi)\right) +1\left(p\right)
+1\left(\rho\right) +4\left(u^{\nu}\right) =18,
\end{equation}
whereas, from (\ref{eq:einstQ}-\ref{eq:CQ}), there exist twenty-five equations. However,
there are nine identities (\ref{eq:id1Q}-\ref{eq:id3Q}) and therefore only
sixteen independent equations yielding an underdetermined system. We can therefore
prescribe two functions out of the eighteen. In case an equation of state is imposed,
\begin{eqnarray}
S(p,\rho)&=&0, \\
\left[\frac{\partial S(p,\rho)}{\partial\rho}
\right]^{2}+\left[\frac{\partial S(p,\rho)}{\partial p}\right]^{2}&>&0, \nonumber
\end{eqnarray}
we can still prescribe one function.

\qquad We can now explore eigenvalues of $T^{\mu}_{\;\nu}$. Assuming that
\begin{eqnarray}
\phi_{,\mu}\not\equiv 0, \label{eq:eigassumpQ} \\
u^{\alpha}s_{\alpha}\equiv 0. \nonumber
\end{eqnarray}
We obtain from (\ref{eq:TQ}-\ref{eq:UQ}) and (\ref{eq:eigassumpQ}) that
\begin{subequations}
\begin{align}
T^{\alpha}_{\;\beta}u^{\beta}=& \left(u^{\beta}\phi_{,\beta}\right) \phi^{;\alpha}
-\left[\rho+ \V +\frac{1}{2}\phi^{;\beta}\phi_{,\beta} \right]u^{\alpha},
\label{eq:evec1Q} \\
T^{\alpha}_{\;\beta}\phi^{;\beta}=& \left[p+ \frac{1}{2} \phi^{;\beta}\phi_{,\beta}
-\V\right]\phi^{;\alpha} +\left[\left(\rho +p \right)u_{\beta}\phi^{;\beta}\right]
u^{\alpha}, \label{eq:evec2Q} \\
T^{\alpha}_{\;\beta}s^{\beta}=&\left[\phi_{,\beta}s^{\beta}\right]\phi^{;\alpha}
-\left[\frac{1}{2} \phi^{;\beta}\phi_{,\beta} +\V -p\right]s^{\alpha}. \label{eq:evec3Q}
\end{align}
\end{subequations}
It is evident from the above equations that, in general, none of the vectors $u^{\beta}$,
$\phi^{;\beta}$, $s^{\beta}$ are eigenvectors. However, consider two special cases. In
case-I, we take the scalar gradient perpendicular to the fluid velocity and $s^{\beta}$:
\begin{eqnarray}
u^{\nu}\phi_{,\nu}&\equiv&0, \label{eq:perpgradQ} \\
s^{\nu}\phi_{,\nu}&\equiv&0. \nonumber
\end{eqnarray}
By (\ref{eq:evec1Q}-\ref{eq:evec2Q}) and (\ref{eq:perpgradQ}), we get:
\begin{subequations}
\begin{align}
T^{\alpha}_{\;\beta}u^{\beta}=& -\left[\rho +\frac{1}{2}\phi^{;\beta}\phi_{,\beta}+ \V\right]u^{\alpha}, \label{eq:class1vecsa} \\
T^{\alpha}_{\;\beta}\phi^{;\beta}=& \left[p  +\frac{1}{2} \phi^{;\beta}\phi_{,\beta}
-\V\right]\phi^{;\alpha}, \label{eq:class1vecsb} \\
T^{\alpha}_{\;\beta}s^{\beta}=&-\left[-p + \frac{1}{2} \phi^{;\beta}\phi_{,\beta} +\V
\right]s^{\alpha}. \label{eq:class1vecsc}
\end{align}
\end{subequations}
In this case, $u^{\beta}$, $\phi^{\beta}$ are eigenvectors and $s^{\beta}$ is a two-fold degenerate eigenvector. The corresponding eigenvalues in (\ref{eq:class1vecsa}-\ref{eq:class1vecsc}) represent the proper mass density and two principal pressures, respectively, of the matter. A positive value of the potential, $\V$, make a positive contribution to the energy density and negative contribution to the pressures. The structure here is algebraically similar to the {\em anisotropic fluid}.

\qquad In case-II, assume that the scalar gradient is colinear with the fluid velocity:
\begin{eqnarray}
\phi^{;\mu}&\equiv& b u^{\mu}, \label{eq:eigassumpQ2} \\
b &\neq& 0,\;\; s^{\nu}\phi_{,\nu}\equiv 0. \nonumber
\end{eqnarray}
By equations (\ref{eq:evec1Q}-\ref{eq:evec3Q}) and (\ref{eq:eigassumpQ2}), we arrive at:
\begin{subequations}
\begin{align}
T^{\alpha}_{\;\beta}u^{;\beta}=& -\left[\rho  -\frac{1}{2}\phi^{;\beta}\phi_{,\beta}+ \V\right]u^{\alpha}, \label{eq:class2vecsa} \\
T^{\alpha}_{\;\beta}\phi^{;\beta}=& -\left[\rho  -\frac{1}{2} \phi^{;\beta}\phi_{,\beta}
+\V\right]\phi^{;\alpha}, \label{eq:class2vecsb} \\
T^{\alpha}_{\;\beta}s^{\beta}=&\left[p-\frac{1}{2} \phi^{;\beta}\phi_{,\beta} -\V
\right]s^{\alpha}. \label{eq:class2vecsc}
\end{align}
\end{subequations}
In this case, $u^{\beta}$, $\phi^{\beta}$ and $s^{\beta}$ are all eigenvectors. The structure here is algebraically similar to a perfect fluid with proper mass density $\rho  -\frac{1}{2}\phi^{;\beta}\phi_{,\beta}+ \V $ and pressure $p-\frac{1}{2} \phi^{;\beta}\phi_{,\beta} -\V $.

\subsection{Tachyonic scalar field}
The tachyon field Lagrangian density is given by:
\begin{equation}
\mathcal{L}_{tach}=-\sqrt{g}\,V(\tach) \left[1+\tach_{,\mu}\tach^{;\mu} \right]^{1/2},
\end{equation}
with $\tach$ representing the tachyon field.

\qquad In the case of the tachyonic scalar field, the governing equations read:
\begin{subequations}
\begin{align}
&\mathcal{E}^{\mu}_{\;\nu}:=G^{\mu}_{\;\nu}-8\pi T^{\mu}_{\;\nu}=0, \label{eq:einstT} \\
&T^{\mu}_{\;\nu}=V(\tach) \left[\frac{\tach^{;\mu}\tach_{,\nu}}{\squart}-\squart \delta^{\mu}_{\;\nu} \right] + \left(\rho+p\right) u^{\mu}u_{\nu} +p\delta^{\mu}_{\;\nu}, \label{eq:TT} \\
&\mathcal{T}_{\nu}:=T^{\mu}_{\;\nu;\mu}=0, \label{eq:consT} \\
&\sigma:=\left[\frac{V(\tach)\, \tach_{,\mu}}{\squart}\right]_{;\mu} -V^{\prime}(\tach)\squart=0, \label{eq:SiT} \\
&\mathcal{U}:=u^{\alpha}u_{\alpha}+1=0, \label{eq:UT} \\
&\mathcal{K}=\left[(\rho +p)u^{\alpha}\right]_{;\alpha}-u^{\alpha} p_{,\alpha}= 0, \label{eq:KT} \\
&\mathcal{F}_{\nu}=(\rho+p)u^{\alpha}u_{\nu;\alpha}+ \left(\delta^{\alpha}_{\;\nu} +u^{\alpha}u_{\nu} \right) p_{,\alpha} =0 ,\label{eq:FT} \\
&\mathcal{C}^{\nu}(g_{\alpha\beta}, g_{\alpha\beta,\gamma})=0. \label{eq:CT}
\end{align}
\end{subequations}
Again the system is underdetermined by two which allows us to prescribe an equation of state and one quantity (usually $\rho$ or $p$). Defining vectors as in (\ref{eq:eigassumpQ}) we find:
\begin{subequations}
\begin{align}
T^{\alpha}_{\;\beta}u^{\beta}=&V(\tach)\frac{u^{\beta}\tach_{,\beta}}{\squart}\tach^{;\alpha}- \left[V(\tach)\squart+\rho\right]u^{\alpha}, \label{eq:evaleqTa} \\
T^{\alpha}_{\;\beta}\tach^{;\beta}=& \left[p-\frac{V(\tach)}{\squart}\right]\tach^{;\alpha} +\left[(\rho+p)u^{\beta}    \tach_{,\beta}\right] u^{\alpha}, \label{eq:evaleqTb}\\
T^{\alpha}_{\;\beta}s^{\beta}=& \frac{V(\tach)}{\squart} s^{\beta}\tach_{,\beta} \tach^{;\alpha} +\left[p-V(\tach)\squart\right]s^{\alpha}. \label{eq:evaleqTc}
\end{align}
\end{subequations}
As with the quintessence field, none of the vectors $u^{\beta}$, $\phi^{;\beta}$, $s^{\beta}$ are eigenvectors. In case the vectors are mutually orthogonal the following equations hold:
\begin{subequations}
\begin{align}
T^{\alpha}_{\;\beta}u^{\beta}=&-\left[\rho +V(\tach)\squart \right]u^{\alpha},\label{eq:evalTaorth} \\
T^{\alpha}_{\;\beta}\tach^{;\beta}=&\left[p-\frac{V(\tach)}{\squart}\right]\tach^{;\alpha}, \label{eq:evalTborth} \\
T^{\alpha}_{\;\beta}s^{\beta}=&\left[p-V(\tach)\squart\right]s^{\alpha}. \label{eq:evalTcorth}
\end{align}
\end{subequations}
Note that in this case, not only are $u^{\beta}$, $\phi^{;\beta}$ and $s^{\beta}$ eigenvectors, but the stress-energy tensor possesses the algebraic structure of an {\em anisotropic fluid}.

\qquad In case the fluid velocity and scalar gradient are colinear, the stress-energy tensor has following structure:
\begin{subequations}
\begin{align}
T^{\alpha}_{\;\beta}u^{\beta}=&-\left[\rho +\frac{V(\tach)}{\squart}\right]u^{\alpha}, \label{eq:evalTa} \\
T^{\alpha}_{\;\beta}\tach^{;\beta}=&-\left[\rho+\frac{V(\tach)}{\squart}\right]\tach^{;\alpha}, \label{eq:evalTb} \\
T^{\alpha}_{\;\beta}s^{\beta}=&\left[p-V(\tach)\squart\right]s^{\alpha}, \label{eq:evalTc}
\end{align}
\end{subequations}
which is similar to that of a {\em perfect fluid}.

\section{The Cauchy problem}
Here we present the Cauchy problem for the self-gravitating system of a perfect fluid and scalar field. 

\begin{figure}[ht!]
\begin{center}
\includegraphics[bb=0 0 280 265, scale=0.6, clip, keepaspectratio=true]{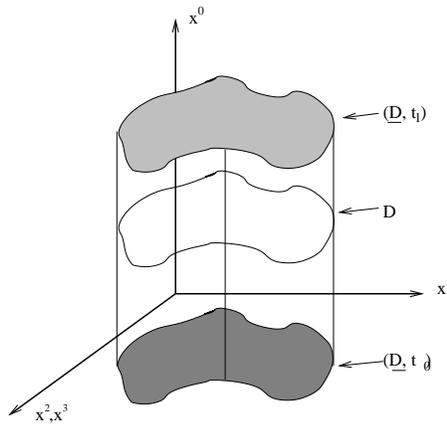}
\caption{{\small Domain for the Cauchy problem. Initial data is specified on the hyper-surface domain $(\underline{D},t_{0})$ and is evolved to the hyper-surface domain $(\underline{D},t_{1})$.}} \label{fig:cauchfig}
\end{center}
\end{figure}

\qquad Let us consider a contractible space-time domain (see figure \ref{fig:cauchfig})
\begin{eqnarray}
D&:=&\underline{D}\times\left(t_{1},t_{0}\right), \label{eq:cauchdomain} \\
x&:=&\left(\mathbf{x},t\right) \in D. \nonumber
\end{eqnarray}

Let a differentiable symmetric tensor field $S_{\mu\nu}(x)$ exist in $D$ of the space-time. We cite Synge's lemma \cite{ref:syngebook} \\
\begin{lemma}{\bf Synge's lemma:} Let $S_{\mu\nu}(x)$ be a symmetric, differentiable tensor field in the domain $D$ of space-time. Then, the following two statements are mathematically equivalent:
\begin{eqnarray}
\mbox{A:}& \;\; &S_{\mu\nu}(x)=0\;\;\mbox{for}\;\; x\in D\, . \nonumber \\
\mbox{B:}& \;\; &S_{ij}(x)-\frac{1}{2}g_{ij}(x)S^{\alpha}_{\;\alpha}(x)=0 \label{eq:synglem1} \\
&&\mbox{and}\;\;S^{\alpha}_{\;\beta;\alpha}=0\;\;\mbox{in}\;\;D,\;\;\mbox{with}\;\;S^{0}_{\;\alpha}(\mathbf{x},t_{0})=0. \nonumber
\end{eqnarray}
\end{lemma}
The above lemma can be applied to Einstein's field equations. In that case, the following two statements are mathematically equaivalent:
\begin{eqnarray}
\mbox{A:}& \;\; &\mathcal{E}_{\mu\nu}(x)=0\;\;\mbox{in}\;\; D\, . \nonumber \\
\mbox{B:}& \;\; &\mathcal{E}_{ij}(x)-\frac{1}{2}g_{ij}(x)\mathcal{E}^{\alpha}_{\;\alpha}(x)=0 \label{eq:einstlem} \\
&&\mbox{and}\;\;\mathcal{T}_{\alpha}=0\;\;\mbox{in}\;\;D,\;\;\mbox{with}\;\;\mathcal{E}^{0}_{\;\alpha}(\mathbf{x},t_{0})=0. \nonumber
\end{eqnarray}
We shall investigate equations (\ref{eq:einstlem}) involving a quintessence scalar field and perfect fluid in the following geodesic normal coordinates:
\begin{eqnarray}
ds^{2}&=&g_{\mu\nu}(x)\,dx^{\mu}dx^{\nu}=-\left(dx^{0}\right)^{2}+g_{ij}(x)\,dx^{i}dx^{j}, \nonumber \\
g^{\sharp}_{ij}(\mathbf{x},t)&:=& g_{ij}(x), \nonumber \\
g^{\sharp}_{ij}(\mathbf{x}, t_{0})&=:& \overline{g}_{ij}(\mathbf{x}), \label{eq:cauchdefs} \\
\Gamma^{i}_{\;jk}&\equiv& \Gamma^{i\sharp}_{\;jk}, \;\;
\Gamma^{i}_{\;0 k}=\frac{1}{2}g^{il \sharp}g^{\sharp}_{lk,0}, \;\;
\Gamma^{0}_{\;ij}=-\frac{1}{2}g^{\sharp}_{ij,0}. \nonumber \\
\overline{g}_{ij,0}(\mathbf{x})&:=&
\left[g^{\sharp}_{ij,0}(\mathbf{x},t)\right]_{|t_{0}},\;\;
\overline{g}_{ij,0,0}(\mathbf{x}):=\left[g^{\sharp}_{ij,0,0}(\mathbf{x},t)\right]_{|t_{0}}
\end{eqnarray}
To clarify, $g^{\sharp}_{ij}$ will subsequently be used to denote the {\em bulk spatial metric} and $\overline{g}_{ij}$ will be used to denote the {\em spatial metric on the initial hyper-surface}.

\qquad Computing the Ricci and Einstein tensor components from (\ref{eq:einstlem}) we obtain:
\begin{subequations}
\begin{align}
R_{ij}(x)=&R^{\sharp}_{ij}(\mathbf{x},t) +\frac{1}{2}g^{\sharp}_{ij,0,0}+\frac{1}{4} g^{\sharp kl}g^{\sharp}_{kl,0}g^{\sharp}_{ij,0} -\frac{1}{2}g^{\sharp kl}g^{\sharp}_{ki,0}g^{\sharp}_{lj,0}, \label{eq:iconst1} \\
G_{i0}(x)=-&\frac{1}{2}g^{\sharp kl}\left[g^{\sharp}_{kl,0;i^{\sharp}}-g^{\sharp}_{li,0;k^{\sharp}}\right], \label{eq:iconst2} \\
G_{00}(x)=& \frac{1}{2} R^{\sharp}(\mathbf{x},t) +\frac{1}{8}
\left[g^{\sharp kl}g^{\sharp}_{kl,0}\right]^{2} -\frac{1}{8}
g^{\sharp km}g^{\sharp ln}g^{\sharp}_{kl,0}g^{\sharp}_{mn,0}. \label{eq:iconst3}
\end{align}
\end{subequations}
Here, sharp symbols ($\sharp$) on indices denotes covariant
differentiation with respect to the three dimensional  metric
$g^{\sharp}_{ij}$. The Ricci and Einstein tensor components
can be expressed as (from (\ref{eq:cauchdefs}) and (\ref{eq:iconst1} -\ref{eq:iconst3})):
\begin{subequations}
\begin{align}
R_{ij}(\mathbf{x},t_{0})=&\overline{R}_{ij}(\mathbf{x}) +\frac{1}{2}\overline{g}_{ij,0,0}(\mathbf{x}) +\frac{1}{4}\overline{g}^{kl} \overline{g}_{kl,0}\overline{g}_{ij,0} -\frac{1}{2} \overline{g}^{kl}\overline{g}_{ki,0}\overline{g}_{lj,0}, \label{eq:newiconst1} \\
G_{i0}(\mathbf{x},t_{0})=-&\frac{1}{2}\overline{g}^{kl}(\mathbf{x}) \left[\overline{g}_{kl,0;\overline{i}} -\overline{g}_{li,0;\overline{k}} \right], \label{eq:newiconst2} \\
G_{00}(\mathbf{x},t_{0})=& \frac{1}{2}\overline{R}(\mathbf{x}) +\frac{1}{8} \left[\overline{g}^{kl}\overline{g}_{kl,0}\right]^{2} -\frac{1}{8} \overline{g}^{km}\overline{g}^{ln}\overline{g}_{kl,0}\overline{g}_{mn,0}. \label{eq:newiconst3}.
\end{align}
\end{subequations}
Note that the \emph{extrinsic curvature components} of the initial hyper-surface are given by $K_{ij}(\mathbf{x})=-\frac{1}{2} \overbar{g}_{ij,0}$ and the above equations are also derivable from the Gauss-Codazzi equations \cite{ref:MTW}, \cite{ref:hawkell}, \cite{ref:stephani}.

\subsection{Quintessence field}
We shall now explore the field equations and subsequent evolution
with quintessence scalar field and perfect fluid sources. Using
Synge's lemma and consequent equations (\ref{eq:einstlem}) and
(\ref{eq:iconst1} - \ref{eq:iconst3}) we arrive at:
\begin{subequations}
\begin{align}
&\mathcal{E}_{ij}-\frac{1}{2}g_{ij}\mathcal{E}^{\alpha}_{\;\alpha}= R^{\sharp}_{ij}(\mathbf{x},t) +\frac{1}{2}g^{\sharp}_{ij,0,0}+\frac{1}{4}g^{\sharp lm}g^{\sharp}_{lm,0}g^{\sharp}_{ij,0} -\frac{1}{2}g^{\sharp lm}g^{\sharp}_{li,0}g^{\sharp}_{mj,0} \nonumber \\
&\;\;\;\;\;\;-8\pi \left\{\phi_{,i}\phi_{,j} + \left(\rho+p\right)u_{i}u_{j} +g^{\sharp}_{ij} \left[\frac{1}{2} g^{\sharp lm}\phi_{,l}\phi_{,m} +\V+\frac{1}{2}(\rho-p) \right]\right\}=0, \label{eq:cauchQE} \\
&\sigma=\frac{1}{\sqrt{g^{\sharp}}} \left[\sqrt{g^{\sharp}}g^{\sharp kl}\phi_{, k}\right]_{,l} -\phi_{,0,0} -\left(\ln \sqrt{g^{\sharp}}\right)_{,0}\phi_{,0} -\vprime=0, \label{eq:cauchQS} \\
&u^{0}=U(u):=\sqrt{1+g^{\sharp mn}u_{m}u_{n}}\, \geq 1, \label{eq:cauchQu} \\
&\mathcal{K}=U(u)\rho_{,0} +u^{l}\rho_{,l} +\frac{(\rho+p)}{\sqrt{g^{\sharp}}} \left\{\left( \sqrt{g^{\sharp}}u^{k}\right)_{,k} +U(u)\left(\sqrt{g^{\sharp}}\right)_{,0} +\sqrt{g^{\sharp}}U_{,0}\right\}=0, \label{eq:cauchQK} \\
&\mathcal{F}_{0}=\left[1-\left(U(u)\right)^{2} \right]p_{,0} -u^{k}U(u)p_{,k} -\left(\rho+p\right) \left\{u^{l}U_{,l} +\frac{1}{2}g^{\sharp bm}g^{\sharp}_{ma,0}u^{a}u_{b}\right\}-UU_{,0}=0, \label{eq:cauchQF} \\
&\mathcal{F}_{j}=(\rho+p) U(u) u_{j,0} -\frac{1}{2}(\rho +p)U
g^{\sharp bm}g^{\sharp}_{mj,0}u_{b} +
\left[p_{,j}+u^{b}u_{j}p_{,b} +Uu_{j}p_{,0} \right]=0. \label{eq:cauchQF2}
\end{align}
\end{subequations}
Moreover, on the initial hyper-surface ($t=t_{0}$), the initial data must satisfy:
\begin{subequations}
\begin{align}
\mathcal{E}_{i0}(\mathbf{x},t_{0})=-&\frac{1}{2}\overline{g}^{mn}(\mathbf{x}) \left[\overline{g}_{mn,0;\overline{i}} -\overbar{g}_{ni,0;\overline{m}}\right] -8\pi \left[\phi_{,i}\phi_{,0}-(\rho +p)U u_{i}\right]_{|t_{0}}, \label{eq:Qinit1} \\
\mathcal{E}_{00}(\mathbf{x},t_{0})=&\frac{1}{2} \overline{R}(\mathbf{x}) +\frac{1}{8} \left[\overbar{g}^{mn}(\mathbf{x}) \overline{g}_{mn,0}\right]^{2} -\frac{1}{8} \overline{g}^{ma}\overline{g}^{nb} \overline{g}_{mn,0} \overline{g}_{ab,0} \nonumber \\
&-8\pi \left\{\frac{1}{2} \left[(\phi_{,0})^{2} +g^{\sharp ab}\phi_{,a}\phi_{,b}\right]+ V(\phi) +(\rho+p) U^{2} -p\right\}_{|t_{0}} =0. \label{eq:Qinit2}
\end{align}
\end{subequations}
The initial data is to be prescribed as:
\begin{subequations}
\begin{align}
\rho(\mathbf{x},t_{0})=&\mu(\mathbf{x}),\;\; p(\mathbf{x},t_{0})= \eta(\mathbf{x}), \label{eq:quintindat1} \\
u_{a}(\mathbf{x},t_{0})=&w_{a}(\mathbf{x}), \label{eq:quintindat2} \\
\phi(\mathbf{x},t_{0})=&\chi(\mathbf{x}), \;\; \phi_{,0}(\mathbf{x},t)_{|t_{0}}=\xi(\mathbf{x}), \label{eq:quintindat3} \\
\overline{g}_{ab}(\mathbf{x})\equiv& g^{\sharp}_{ab}(\mathbf{x},t_{0})=\gamma_{ab}(\mathbf{x}), \label{eq:quintindat4} \\
g_{ab,0}^{\sharp}(\mathbf{x},t)_{|t_{0}}=&\psi_{ab}(\mathbf{x}).
\label{eq:quintindat5}
\end{align}
\end{subequations}
The given functions $\mu$, $\sigma$, $w_{a}$ and $\xi$ are of class $C^{1}$ in $\underline{D} \subset \mathbb{R}^{3}$, the functions $\psi_{ab}$ are of class $C^{2}$ and the functions $\gamma_{ab}$ are of class $C^{3}$. Moreover, the consistency conditions (\ref{eq:Qinit1} - \ref{eq:Qinit2}) must be satisfied as:
\begin{subequations}
\begin{align}
&\frac{1}{2}\gamma^{mn} \left[\psi_{mn;\overline{a}}-\psi_{na;\overline{m}} \right] +8\pi \left\{ \xi\chi_{,a} -\left[\mu+\eta\right]U\,w_{a}\right]=0, \label{eq:Qconsist1} \\
&\frac{1}{2}\overline{R}+\frac{1}{8} \left[\gamma^{mn}
\psi_{mn}\right]^{2} -\frac{1}{8} \gamma^{ma} \gamma^{nb}
\psi_{mn}\psi_{ab} \nonumber \\
& \;\;\; -8\pi \left[\frac{1}{2} \left(\xi^{2}
+\overline{\gamma}^{ab} \chi_{,a}\chi_{,b} \right) +V(\chi)
+\left(\mu+\eta\right) U^{2} -\eta \right]=0. \label{eq:Qconsist2}
\end{align}
\end{subequations}
(Here, barred quantities are derived from the prescribed metric $\gamma_{mn}(\mathbf{x})\;$). The system of three dimensional partial differential equations is underdetermined so that infinitely many solutions should exist locally. Restricting the field equations to the initial hyper-surface, $t=t_{0}$, we can rearrange the equations symbolically as:
\begin{eqnarray}
\frac{1}{2}\overline{g}_{ij,0,0}&=&\mbox{initial data and their spatial derivatives}, \nonumber \\
\phi_{,0,0\; |t_{0}}&=& \;\;\; " \;\;\;\;\;\;\;\; " \;\;\;\;\;\; " \;\;\;\;\;\; " \;\;\;\;\;\;\;\;\; " \;\;\;\;\;\;\;\;\; " \;\;\;\;\;\;\;\;\;\;, \nonumber \\
U\rho_{,0\;|t_{0}}&=& \;\;\; " \;\;\;\;\;\;\;\; " \;\;\;\;\;\; " \;\;\;\;\;\; " \;\;\;\;\;\;\;\;\; " \;\;\;\;\;\;\;\;\; " \;\;\;\;\;\;\;\;\;\;,\nonumber \\
\left\{\left[1-U^{2}\right]p_{,0}\right\}_{|t_{0}}&=& \;\;\; " \;\;\;\;\;\;\;\; " \;\;\;\;\;\; " \;\;\;\;\;\; " \;\;\;\;\;\;\;\;\; " \;\;\;\;\;\;\;\;\; " \;\;\;\;\;\;\;\;\;\;,\nonumber \\
\left[\mu+\eta\right]U\,u_{a,0\;|t_{0}}&=&\;\;\; " \;\;\;\;\;\;\;\; " \;\;\;\;\;\; " \;\;\;\;\;\; " \;\;\;\;\;\;\;\;\; " \;\;\;\;\;\;\;\;\; " \;\;\;\;\;\;\;\;\;\; .\nonumber
\end{eqnarray}
Thus, the higher derivatives of $g_{ij}$, $\phi$, $\rho$, $p$, and $u_{a}$ at the initial hyper-surface are determined by allowable Cauchy data and their spatial derivatives.  Assuming the functions $g_{ij}(\mathbf{x},t)$, $\phi(\mathbf{x},t)$, $\rho(\mathbf{x},t)$, $p(\mathbf{x},t)$, and $u_{a}(\mathbf{x},t)$ are \emph{real-analytic functions} in $t_{0}\leq t <t$, we can determine arbitrary higher order derivatives with respect to $x^{0}$ by differentiating the field equations in (\ref{eq:cauchQE} - \ref{eq:cauchQF2}) and restricting subsequently those equations to the initial hyper-surface. Thus, the power series:
\begin{subequations}
\begin{align}
g_{ij}(\mathbf{x},t)=&\gamma_{ij}(\mathbf{x})+ \psi_{ij}(\mathbf{x})(t-t_{0}) +\frac{1}{2} \overline{g}_{ij,0,0} (t-t_{0})^{2} +... \;\;, \label{eq:gevolve} \\
\phi(\mathbf{x},t)=& \chi(\mathbf{x}) +\xi(\mathbf{x})(t-t_{0}) +\frac{1}{2} \phi_{,0,0\;|t_{0}}(t-t_{0})^{2} + ...\;\;, \label{eq:phievolve} \\
\rho(\mathbf{x},t)=& \mu(\mathbf{x}) +\rho_{,0\;|t_{0}}(t-t_{0}) + ... \;\; ,  \label{eq:rhoevolve} \\
p(\mathbf{x},t)=& \eta(\mathbf{x}) +p_{,0\;|t_{0}}(t-t_{0})+ ... \;\;, \label{eq:pevolve} \\
u_{a}(\mathbf{x},t)=& w_{a}(\mathbf{x}) +u_{a,0\;|t_{0}} (t-t_{0}) +...\;\; , \label{eq:uevolve}
\end{align}
\end{subequations}
can be generated. By the real analyticity conditions, there exist a $t_{1}>t_{0}$ such that all the power series in (\ref{eq:gevolve}-\ref{eq:uevolve}) converge absolutely for all $t\in(t_{1},t_{0}]$ and converge uniformly for all $t\in \left[t_{1}-\delta, t_{0}\right]$ for a sufficiently small $\delta > 0$.

\subsection{Tachyonic field}
In the case of the tachyonic scalar field, the field and
supplemental equations may be expressed as:
\begin{subequations}
\begin{align}
&\mathcal{E}_{ij}- \frac{1}{2}g_{ij} \mathcal{E}^{\mu}_{\;\mu}=
R^{\sharp}_{ij}+ \frac{1}{2} g^{\sharp}_{ij,0,0}
+\frac{1}{4}g^{\sharp mn}g^{\sharp}_{mn,0}g^{\sharp}_{ij,0}
-\frac{1}{2} g^{\sharp
mn}g^{\sharp}_{mi,0}g^{\sharp}_{nj,0} \nonumber \\
&\;\;\;-8\pi \left\{\V \left[\frac{\tach_{,i}\tach_{,j}
+g^{\sharp}_{ij} \left(1+ \frac{1}{2} \tach^{;\mu} \tach_{,\mu}
\right)}{\squart} \right]+ (\rho+p)u_{i}u_{j}
+\frac{1}{2}g^{\sharp}_{ij}(\rho -p) \right\}=0, \label{eq:Etach} \\
&\sigma=\frac{1}{\sqrt{g^{\sharp}}} \left[\sqrt{g^{\sharp}}
g^{\sharp ab}\tach_{,a} \right]_{,b} -\tach_{,0,0}
-\left(\ln \sqrt{g^{\sharp}}\right)_{,0} \tach_{,0} -\left[\ln |\V | \right]^{\prime}  \nonumber \\
&\;\;\;\;\;\; -\left[1+\tach^{;\mu}\tach_{,\mu} \right]^{-1} \left[\tach_{,i} \tach_{,j} g^{\sharp \,nj}g^{\sharp\, bi} \left(\tach_{,n;b^{\sharp}} +\frac{1}{2}g^{\sharp}_{bn,0}\tach_{,0}\right) \left(\tach_{,0}\right)^{2}\tach_{,0,0} \right. \nonumber \\
&\;\;\;\;\;\; \left. -2\tach_{,0}\tach_{,j}g^{\sharp\,nj}\tach_{,n;0}\right]
= 0, \label{eq:sigtach} \\
& u^{0}=:U:=\sqrt{1+g^{\sharp ab}u_{a}u_{b}} \geq 1, \label{eq:utach}  \\
&\mathcal{K}:=U\,\rho_{,0} +u^{a} \rho_{,a}
+\frac{(\rho+p)}{\sqrt{g^{\sharp}}}
\left[\left(\sqrt{g^{\sharp}}u^{a}\right)_{,a}+
U\sqrt{g^{\sharp}}_{,0} + \sqrt{g^{\sharp}} U_{,0} \right]=0, \label{eq:Ktach} \\
&\mathcal{F}_{0}=\left[1-\left(U\right)^{2} \right]p_{,0}
-u^{k}U(u)p_{,k} -\left(\rho+p\right)
\left\{u^{l}U_{,l} +\frac{1}{2}g^{\sharp bm}g^{\sharp}_{ma,0}u_{b}\right\}-UU_{,0}=0, \label{eq:F0tach}  \\
&\mathcal{F}_{j}=(\rho+p) U(u) u_{j,0} -\frac{1}{2}(\rho +p)U
g^{\sharp bm}g^{\sharp}_{mj,0}u_{b} +
\left[p_{,j}+u^{b}u_{j}p_{,b} +Uu_{j}p_{,0} \right]=0. \label{eq:F1tach}
\end{align}
\end{subequations}
As before, there exist constraints on the initial data on the
$t=t_{0}$ hyper-surface:
\begin{subequations}
\begin{align}
\mathcal{E}_{a0}(\mathbf{x},t_{0})=&-\frac{1}{2}\overline{g}^{mn}
\left[\overline{g}_{mn,0;\overline{a}}-\overline{g}_{na,0;\overbar{m}}
\right] \nonumber \\
&- 8\pi \left\{\V \left[\frac{\tach_{,a}\tach_{,0}}{\sqrt{1+
g^{\sharp rs}\tach_{,r}\tach_{,s}-(\tach_{,0})^{2}}}\right]
-(\rho + p)U
u_{a} \right\}_{|t_{0}}=0, \label{eq:tachcons1} \\
\mathcal{E}_{00}(\mathbf{x},t_{0})=& \frac{1}{2}
\overline{R}+\frac{1}{8} \left[\overline{g}^{mn}
\overline{g}_{mn,0} \right]^{2} - \frac{1}{8} \overline{g}^{ma}
\overline{g}^{nb} \overline{g}_{mn,0} \overline{g}_{ab,0}
\nonumber \\
&-8\pi \left\{\V\left[\frac{1+g^{\sharp ab}\tach_{,a}
\tach_{,b}}{\sqrt{1+g^{\sharp mn} \tach_{,m}\tach_{,n}
-(\tach_{,0})^{2}}} \right]+ (\rho+p) U^{2} - p\right\}_{|t_{0}}
=0. \label{eq:tachcons2}
\end{align}
\end{subequations}
The initial data is prescribed as in the quintessence case
(\ref{eq:quintindat1})-(\ref{eq:quintindat5}) with $\phi$ replaced
by $\tach$ (differentiability requirements remain the same). In
this case, the consistency equations
(\ref{eq:tachcons1})-(\ref{eq:tachcons2}) become:
\begin{subequations}
\begin{align}
&\frac{1}{2} \gamma^{mn} \left[\psi_{mn;\overline{a}} -
\psi_{na;\overline{m}}\right] +8\pi \left\{V(\chi)
\left[\frac{\xi\chi_{,a}}{\sqrt{1+\gamma^{rs}\chi_{,r}\chi_{,s}
-\xi^{2}}} \right] -\left(\mu+ \eta\right)U\,w_{a} \right\}=0, \label{eq:inittacha} \\
&-\frac{1}{2}\overline{R} -\frac{1}{8} \left[\gamma^{mn}
\psi_{mn}\right]^{2} +\frac{1}{8} \gamma^{ma} \gamma^{nb}
\psi_{mn} \psi_{ab} \nonumber \\
&\;\;\; +8\pi \left\{V(\chi)
\left[\frac{1+\gamma^{ab}\chi_{,a}\chi_{,b}}{\sqrt{1+
\gamma^{mn}\chi_{,m}\chi_{,n} -\xi^{2}}}\right] +(\mu+\eta) U^{2}
-\eta\right\} =0. \label{eq:inittachb}
\end{align}
\end{subequations}
At this point, the evolution is governed by the equations (\ref{eq:gevolve}) - (\ref{eq:uevolve}) as before.

\subsection{A specific example}
It is instructive to demonstrate how the above scheme works with
an explicit example. We shall study the numerical evolution of a
system whose analytic properties are known. Comparison with a known solution will show whether or not the method works as well as provide a benchmark on its convergence properties. Specifically, we
consider the evolution of a constant scalar field in an otherwise
empty, flat Friedmann-Lema\'{i}tre-Robertson-Walker (FLRW) space-time.
Although extremely simple, this example is pedagogically
useful as it serves well to elucidate the employment of the
scheme without unnecessary complications which arise from more
complex systems. As well, the constant field evolution is identical for both the quintessence and tachyonic scenario. We show how to extract various quantities for
the Cauchy evolution and compare with the known analytic result.

\qquad Since the source consists of only the constant scalar
field, we can immediately set the following initial data:
\begin{equation}
\chi(\mathbf{x})=\chi_{0}=\mbox{const},\;\; \xi(\mathbf{x})=
\mu(\mathbf{x})= \eta(\mathbf{x})=w_{a}(\mathbf{x})\equiv 0.
 \label{eq:exinitdat}
\end{equation}
With this prescription, the equation pairs (\ref{eq:Qconsist1} - \ref{eq:Qconsist2}) and (\ref{eq:inittacha} - \ref{eq:inittachb}) both take the form:
\begin{subequations}
\begin{align}
& \frac{1}{2}\gamma^{mn}\left[\psi_{mn;\overline{a}}-\psi_{na;\overbar{m}} \right] = 0, \label{eq:exampconst1} \\
& -\frac{1}{2}\overline{R} -\frac{1}{8}\left[\gamma^{mn}\psi_{mn}\right]^{2} +\frac{1}{8} \gamma^{ma} \gamma^{nb} \psi_{mn}\psi_{ab} +8\pi V(\chi_{0}) =0. \label{eq:exampconst2}
\end{align}
\end{subequations}
It should be noted that in general, the consistency equations, although underdetermined, are difficult to solve even in vacuum and one must appeal to numerical techniques for solutions.

\qquad As well, we consider the spatially flat FLRW metric:
\begin{equation}
\left[\gamma_{ab}(\mathbf{x})\right]=a^{2}(t_{0})\left[
\begin {array}{ccc}
1&0&0\\
\noalign{\medskip}
0&r^{2}&0\\
\noalign{\medskip}
0&0&r^{2}\sin^{2}\theta
\end {array}
\right], \nonumber
\end{equation}
with $a(t_{0})$ the value of the scale factor on the initial data
hyper-surface. We consider an inflationary scenario and prescribe:
\begin{align}
a^{2}(t_{0})=&e^{2t_{0}}, \label{eq:exampa}  \\
[a^{2}(t)]_{,0|t=t_{0}}=&2e^{2t_{0}}. \nonumber
\end{align}
At this point, all the allowable initial data has been
prescribed. The Einstein-scalar field equations
(\ref{eq:Etach}-\ref{eq:sigtach}) dictate that
\begin{align}
\chi(\mathbf{x})=& \chi_{0}, \nonumber \\
V(\chi_{0})=&\frac{3}{8\pi}, \label{eq:exampv} \\
\vprime\equiv& 0. \nonumber
\end{align}
At this point it is useful to reiterate that the above
prescription is only valid \emph{if} it satisfies the constraint
equations (\ref{eq:tachcons1} - \ref{eq:tachcons2}). In this example, the relevant
values for the constraint equations are:
\begin{align}
\gamma_{ab}(\mathbf{x})=&e^{2t_{0}}\delta_{ab}, \nonumber \\
\psi_{mn}=2e^{2t_{0}} \delta_{mn} \label{eq:exampinit} \\
\psi_{ab\; ;\overline{m}}(\mathbf{x})\equiv & 0, \nonumber \\
\overline{R} \equiv& 0. \nonumber\
\end{align}
It is a simple matter to check that these quantities indeed
satisfy the constraint equations (\ref{eq:exampconst1}) and (\ref{eq:exampconst2}).

\begin{figure}[ht!]
\begin{center}
\includegraphics[bb=89 416 557 678, scale=0.6, clip, keepaspectratio=true]{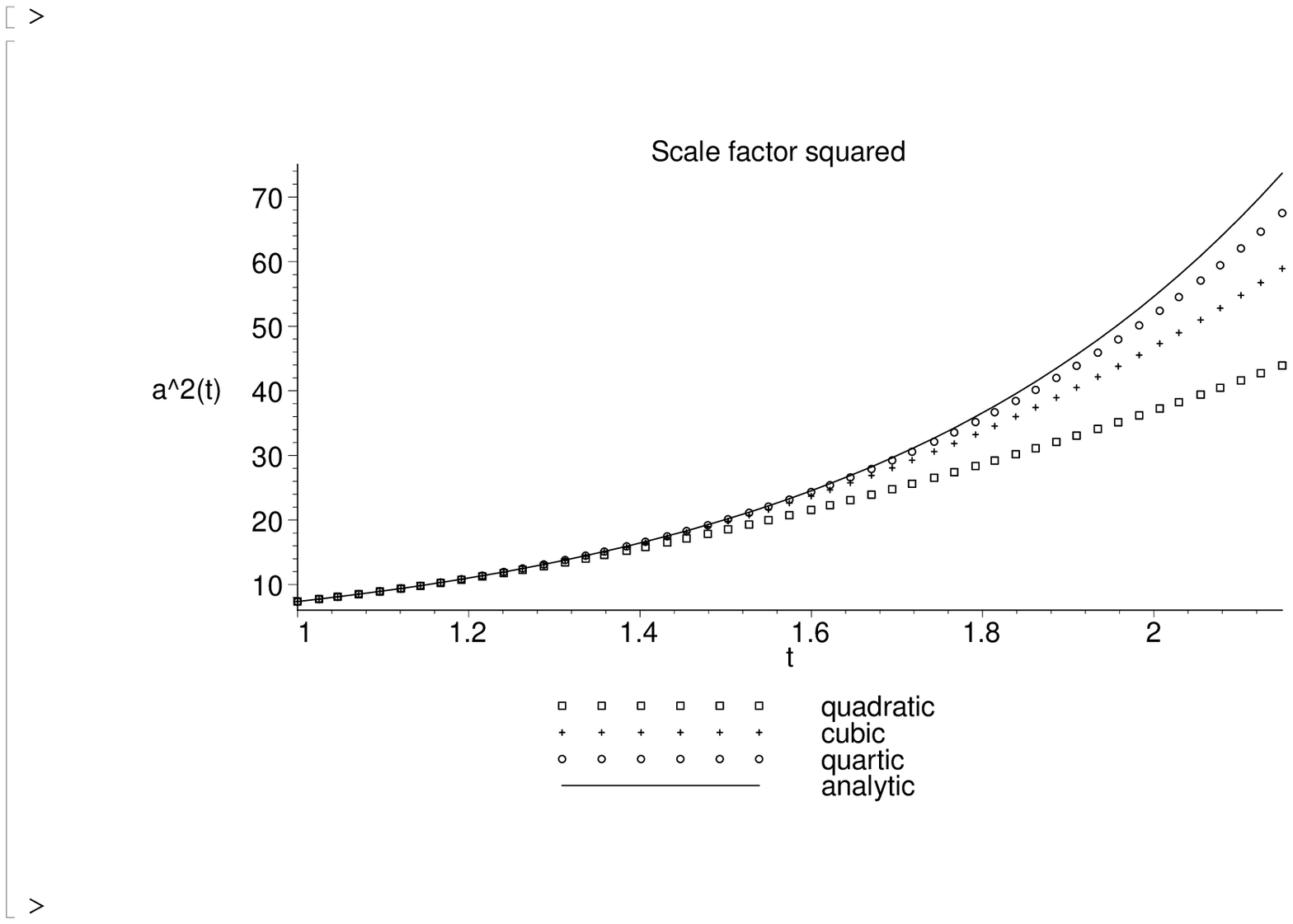}
\caption{{\small Comparison of the square of the cosmological scale factor. The dotted lines represent the numerically evolved Cauchy data utilising the scheme outlined in this paper to various orders in $t-t_{0}$ (quadratic, cubic, quartic). The solid line represents the analytic result $(e^{2t})$.}} \label{fig:evolved}
\end{center}
\end{figure}

\qquad We show both the Cauchy evolved and the analytic $(e^{2t})$ square of the scale factor in figure \ref{fig:evolved}. As expected, there is excellent agreement at small values of $\Delta t$. Deviations at larger $\Delta t$ may be minimized by retaining more terms in the Taylor series (\ref{eq:gevolve}) as can be seen from the vatious dotted lines in the figure (see figure caption). It may readily be verified that the other parameters to be evolved will display the proper evolution by a simple inspection of (\ref{eq:gevolve} - \ref{eq:uevolve}). That is, in this simple example, all other parameters will retain their initial values as prescribed by (\ref{eq:exinitdat}).

\section{Concluding remarks}
Both the quintessence and tachyonic scalar field, supplemented with a perfect fluid, were considered in the context of general relativity. The general mathematical properties of the system, including the eigenvalue structure were studied. It is seen that the system may behave either as a two component perfect fluid or an anisotopic fluid, the anisotropy being due to the properties of the scalar field. Finally, in the geodesic coordinates, the Cauchy problem as well as the initial constraint equations have been derived. Assuming analyticity, the Cauchy scheme presented here is convergent. The scheme is iterative and may easily be executed by computer.

\linespread{0.6}
\bibliographystyle{unsrt}

\end{document}